\documentclass[aps,prd,preprint,superscriptaddress]{revtex4}
\usepackage{amsmath}
\usepackage{amssymb}
\usepackage{epsfig}
\usepackage{ulem}
\usepackage{graphics}
\usepackage{indentfirst}
\usepackage{color}
\usepackage{multirow}

\newcommand{\pb}{\ensuremath{\bar{\Phi}}}

\newcommand{\te}{\ensuremath{\theta}}
\newcounter{RomanNumber}
\newcommand{\MyRoman}[1]{\setcounter{RomanNumber}{#1}\Roman{RomanNumber}}
\newcommand{\kah}{K\"ahler }
\newcommand{\wb}{\ensuremath{\bar{W}}}
\newcommand{\xb}{\ensuremath{\bar{X}}}

\def\be{\begin{equation}}
\def\ee{\end{equation}}
\def\bea{\begin{eqnarray}}
\def\eea{\end{eqnarray}}

\begin{document}

\preprint{ACT-06-15, MI-TH-1527}

\vspace*{-0.4cm}
\title{Helical Phase Inflation via Non-Geometric Flux Compactifications: from Natural to Starobinsky-like Inflation}

\vspace*{-0.4cm}

\author{Tianjun Li} 
\affiliation{{\scriptsize State Key Laboratory of Theoretical Physics
and Kavli Institute for Theoretical Physics China (KITPC),
      Institute of Theoretical Physics, Chinese Academy of Sciences,
Beijing 100190, P. R. China}}

\affiliation{{\scriptsize School of Physical Electronics,
University of Electronic Science and Technology of China,
Chengdu 610054, P. R. China}}

\author{Zhijin Li}

\affiliation{{\scriptsize George P. and Cynthia W. Mitchell Institute for
Fundamental Physics and Astronomy,
Texas A\&M University, College Station, TX 77843, USA}}

\author{Dimitri V. Nanopoulos}

\affiliation{{\scriptsize George P. and Cynthia W. Mitchell Institute for
Fundamental Physics and Astronomy,
Texas A\&M University, College Station, TX 77843, USA}}

\affiliation{{\scriptsize Astroparticle Physics Group, Houston Advanced Research Center (HARC), Mitchell Campus, Woodlands, TX 77381, USA}}

\affiliation{{\scriptsize Academy of Athens, Division of Natural Sciences, Athens 10679, Greece} \vspace*{1.2 cm}}

\begin{abstract}
 {\small  We show that a new class of helical phase inflation models can be simply realized in minimal
supergravity, wherein the inflaton is the phase component of a complex
field and its potential admits a deformed helicoid structure. We find a new unique complex-valued
 index $\chi$ that characterizes almost the entire region of the $n_s-r$ plane favored by new Planck
observations.  Continuously varying the index $\chi$, predictions
interpolate from quadratic/natural inflation parameterized by a phase/axion decay constant to
Starobinsky-like inflation parameterized by the $\alpha$-parameter.  We demonstrate that the simple
supergravity construction realizing Starobinsky-like inflation can be obtained from a more microscopic
model by integrating out heavy fields, and that the flat phase direction for slow-roll inflation is protected
by a mildly broken global $U(1)$ symmetry. 
 We study the geometrical origin of the index $\chi$, and find that it corresponds to a linear constraint
relating \kah moduli.  We argue that such a linear constraint is a natural result of moduli
stabilization in Type \MyRoman{2} orientifold compactifications on Calabi-Yau threefolds
with geometric and non-geometric fluxes. Possible choices for the index $\chi$ are discrete points
on the complex plane that relate to the distribution of supersymmetric Minkowski vacua on moduli space.
More precise observations of the inflationary epoch in the future may provide a better estimation of the index
 $\chi$. Since $\chi$ is determined by the fluxes and vacuum expectation values of
complex structure moduli, such observations would characterize the geometry of the internal space as well.}
\end{abstract}


\maketitle

\section{Introduction}

Inflation \cite{Staro, oldinf} has attracted widespread attention in the past few decades. The
inflationary epoch is crucial for the cosmic evolution, and provides a unique opportunity to
probe physics close to the grand unification scale, far beyond the
scope directly accessible in the laboratory.  A principle challenge to the construction of inflationary
models within the $N=1$ supergravity or superstring theories is the so-called $\eta$ problem.  Specifically,
the inflaton potential obtained in these contexts is usually too steep to trigger slow-roll inflation.
Moreover, it requires trans-Planckian field excursion for the generation of sizable tensor fluctuations
\cite{Lyth:1996im}, rendering Planck-suppressed higher-dimensional operators non-negligible.

Helical phase inflation \cite{Li:2014vpa, Li:2014unh} was proposed as a solution to both the
$\eta$ problem$^1$~\footnotetext[1]{The $\eta$ problem can also be solved by Heisenberg symmetry in
no-scale supergravity \cite{Cremmer:1983bf, nsp, Gaillard:1995az, Antusch:2008pn} or shift symmetry in minimal
supergravity \cite{Kawasaki:2000yn}. Supergravity inflation with broken shift or global
$U(1)$ symmetries was studied in \cite{Li:2013nfa, Harigaya:2014qza, Li:2015mwa}.} and
the trans-Planckian field excursion problem. In helical phase inflation the inflaton is a
pseudo-Nambu-Goldstone boson (PNGB), the phase component of a complex field. A PNGB was first employed
as the inflaton in \cite{Freese:1990rb} in order to protect the flat potential against quantum loop corrections.
The potential of a complex field admits helicoid structure, and during inflation its radial component is
strongly stabilized, while the inflaton evolves along a local valley, tracing a beautiful
helical trajectory.  In this model, the $\eta$ problem is automatically solved by the global $U(1)$ symmetry
of the minimal \kah potential $K=\Phi\pb$.  This $U(1)$ symmetry is broken in the holomorphic superpotential and
leads to phase monodromy.  Phase rotation provides a proxy for trans-Planckian field excursion, whereas
the ``physical'' field does not evolve into the super-Planckian domain, where quantum gravity
effects are likely to break slow-roll conditions.
As argued in \cite{Li:2014unh}, such supergravity constructions are necessarily effective
descriptions of a more fundamental theory with heavy fields integrated out.

In helical phase inflation, the helical trajectory and phase monodromy of the superpotential are similar to
the axion monodromy inflation scenario, as realized via the DBI action of wrapped D5-branes in Ref.~\cite{McAllister:2008hb}.
Likewise, the axion alignment mechanism was suggested in Ref.~\cite{Kim:2004rp},
in order to obtain super-Planckian axion decay constant, and was investigated as a new type of
axion monodromy in \cite{Choi:2014rja, Tye:2014tja, Kappl:2014lra}.
Inflaton dynamics in helical phase inflation associated with explicit breaking of the global $U(1)$ symmetry
may be similarly dubbed phase-axion alignment.
Additionally, a PNGB has been employed as the inflaton in recent studies on inflation models with stabilized or
almost stabilized radial component \cite{McDonald:2014oza, McDonald:2014nqa, Barenboim:2014vea,
McDonald:2014rha, Carone:2014cta, Barenboim:2015zka,
Li:2015mwa, Achucarro:2015rfa, Blanco-Pillado:2015bha, Barenboim:2015lla}.  However, most of these models
do require super-Planckian field excursion during inflation and predict large tensor fluctuation
with a tensor-to-scalar ratio $r\sim0.1$~$^2$\footnotetext[2]{It is shown in \cite{Peloso:2015dsa} that small tensor-to-scalar ratio can be generated in aligned natural inflation initiated close to a saddle point.}.
Recent Planck observations on cosmic microwave background \cite{Planck:2015xua,
Ade:2015oja}, as well as a joint analysis utilizing B-mode polarization data from the BICEP2/Keck Array
\cite{Ade:2015tva}, have provided tighter constraints on inflationary observables, particularly the
tensor-to-scalar ratio, which is $r<0.08$ at the $95\%$ confidence level. As a consequence, large field inflation
models with power-law potentials $V(\phi)\propto\phi^n$ have been ruled out for $n\geqslant2$, and
natural inflation is now also in tension with current data.  In contrast, the Starobinsky model \cite{Staro}
predicts a small tensor-to-scalar ratio $r\simeq 0.003$, which remains entirely consistent with
the latest observations.

Since the discovery of no-scale supergravity realizations of the Starobinsky model
\cite{Ellis:2013xoa, Ellis:2013nxa}, numerous generalization and extensions of the idea
have been proposed.  Specifically, by introducing one additional parameter, the Starobinsky model can
interpolate to quadratic or natural inflation~\cite{Ellis:2013nxa, Ferrara:2013rsa,
Kallosh:2013yoa, Ellis:2014cma, Ellis:2014gxa, Kounnas:2014gda, Ellis:2014opa, Higaki:2015kta, Kannike:2015apa, Ozkan:2015iva, Ellis:2015xna}.
The problem of trans-Planckian field excursion has been carefully considered in quadratic and natural
inflation, while it usually is ignored for Starobinsky-like inflation, since the tensor-to-scalar ratio is
lower than the Lyth bound $r\sim0.01$.
For Starobinsky-like inflation with a potential
\be
V(\phi)=M^4(1-e^{-\alpha\phi})^2,
\ee
the tensor-to-scalar ratio is given by $r=\frac{8}{\alpha^2N^2}$, with an e-folding number $N\in[50,
60]$. For small $\alpha\leqslant0.5$, the tensor-to-scalar ratio
is above the Lyth bound, and the model approximates quadratic inflation. For larger $\alpha$,
the field excursion during inflation can be expressed as
\be
\Delta\phi\approx\frac{1}{\alpha}\log(2\alpha^2N),
\ee
in Planck units ($M_P=1$), which reduces to $\Delta\phi\approx5$ for typical parameter values
$\alpha=\sqrt{\frac{2}{3}}$ and $r\approx0.003$.  Therefore, the Starobinsky-like inflation scenario
is indeed subject to trans-Planckian field excursions, even though the tensor-to-scalar ratio is
below the Lyth bound.  In order to avoid higher order corrections from quantum gravity effects, which
are important in the super-Planckian regime and are likely to violate the slow-roll criteria, related
models of inflation must be studied in the context of a UV-completion, such as string theory.
Starobinsky-like inflation with a string-theoretic embedding has
been studied in Refs.~\cite{Cicoli:2008gp, Burgess:2013sla, Cicoli:2013oba, Blumenhagen:2015qda}.
Another interesting solution is the realization of Starobinsky-like inflation within the
sub-Planckian region, while allowing trans-Planckian field excursions to be undertaken by an ``unphysical''
degree of freedom, such as the phase of a complex field which does not admit any polynomial higher order corrections.

In this work, we show that Starobinsky-like inflation can be simply realized based on the supergravity
setup for helical phase inflation \cite{Li:2014unh}.
Actually, the supergravity setup for Starobinsky-like helical phase inflation is the same as for natural
inflation, except that the latter case features a real-valued superpotential parameter that is pure imaginary in the former case.
Admitting a complex-valued phase for this parameter,
predictions for $n_s-r$ thereby interpolate between natural inflation and Starobinsky-like
inflation, and regions of the $n_s-r$ plane favored by new Planck observations may be characterized by a
single complex index $\chi$.  The supergravity model is expected to be obtained from a more microscopic
model after integrating out heavy fields.  In particular, we find that the index $\chi$ has an interesting
geometrical origin associated with non-geometric flux compactification.

Non-geometric fluxes are motivated from T-duality between the Type \MyRoman{2}A and \MyRoman{2}B
string theories \cite{Kachru:2002sk, Shelton:2005cf}. In the low-energy effective $N=1$ supergravity
theory obtained from type \MyRoman{2} string compactification,
T-duality is preserved in the action for RR fluxes while this is not the case for NSNS fluxes,
leading to the expectation of new fluxes that are T-dual to the NSNS variety. The
geometric flux arises from the NSNS flux by invoking T-duality along a direction of the internal space,
and it relates to compactification on a twisted torus.
By taking T-duality along extra internal directions, one obtains $Q$ or $R$ type fluxes without clear
geometric explanation.  Geometric and non-geometric fluxes introduce coupling terms in the superpotential
for \kah moduli and uplift these directions at the perturbative level so that they can play important roles in
string phenomenology. Moduli stabilization and supersymmetric Minkowski vacua based on non-geometric
flux compactification have been studied extensively in
Refs.~\cite{Aldazabal:2006up, Shelton:2006fd,Micu:2007rd, Font:2008vd, Guarino:2008ik,
deCarlos:2009qm, Aldazabal:2011yz, Hassler:2014mla, Blumenhagen:2015kja}.
It should be noted that, distinct from NSNS and RR fluxes, compactifications with non-geometric fluxes
suffer from the dilution problem.  After turning on non-geometric fluxes, back-reaction on
the internal metric can not be treated by taking a large volume limit with diluted fluxes.
The four-dimensional vacua with non-geometric fluxes have been uplifted to ten dimensions based on the $\beta$-supergravity framework
\cite{Andriot:2014qla}.
The effective supergravity action from non-geometric fluxes is expected to partially reflect the dynamics around vacua of full string theory.

This paper is organized as follows. In Section \MyRoman{2}, we study Starobinsky-like helical phase
inflation and compare its predictions with new Planck observations, showing that the
complex-valued index $\chi$ can characterize regions in the $n_s-r$ plane preferred by new Planck data.
In Section \MyRoman{3} we study a more fundamental realization of helical phase inflation based on both
perturbative and non-perturbative effects. The phase monodromy is identified as a global $U(1)$
symmetry mildly breaking at the inflation energy scale. In Section \MyRoman{4} we study the geometrical
origin of the index $\chi$ in type \MyRoman{2} orientifold compactifications with geometric and
non-geometric fluxes and show that the index $\chi$ is determined by the flux quanta and vacuum
expectation values of complex structure moduli. Conclusions are given in Section \MyRoman{5}.

\section{Starobinsky-like Helical Phase Inflation}

The $N=1$ minimal supergravity setup for Starobinsky-like helical phase inflation is rather simple, with the
\kah potential and superpotential given as follows$^3$~\footnotetext[3]{In this model two superfields are employed,
recently it was shown in \cite{Ketov:2015tpa} that the helical phase inflation can also be realized with only one superfield.}:
\be
K=\Phi\pb+X\bar{X}-g(X\bar{X})^2, ~~~~W=a\frac{X}{\Phi}(\Phi^{ic}-1). \label{set1}
\ee
A similar supergravity model was proposed in \cite{Li:2014unh} for natural inflation, wherein
the superpotential contains a real parameter in the term $\Phi^b, b\ll1$ instead of $\Phi^{ic}$. Here, the
imaginary exponent of $\Phi$ seems to be unusual at first glance, although we will show that it has a clear
geometrical origin associated with non-geometrical flux compactification of type \MyRoman{2}B superstring
theory. There is a global $U(1)$ symmetry in the \kah potential, shifts in which introduces
phase monodromy in the superpotential $W$:
\be
\Phi\rightarrow\Phi e^{2\pi i}, ~~~~~~  K\rightarrow K,  ~~~~~~   W\rightarrow W+a\frac{X}{\Phi}\Phi^{ic}(e^{-2\pi c}-1). \label{mon1}
\ee
By employing the phase of $\Phi$ as an inflaton, the well-known $\eta$ problem for supergravity inflation is
absent, since the \kah potential is phase independent. The phase monodromy in (\ref{mon1}) never
reverts to the original $W$, and so is different from that associated with natural inflation \cite{Li:2014unh}, wherein the
superpotential is cyclically restored after a sufficient long phase rotation ($\Delta\te>2\pi$ with
super-Planckian phase/axion decay constant). Similar differences also exist between the respective inflaton potentials.

\begin{figure}
\centering
\includegraphics[width=120mm, height=100mm,angle=0]{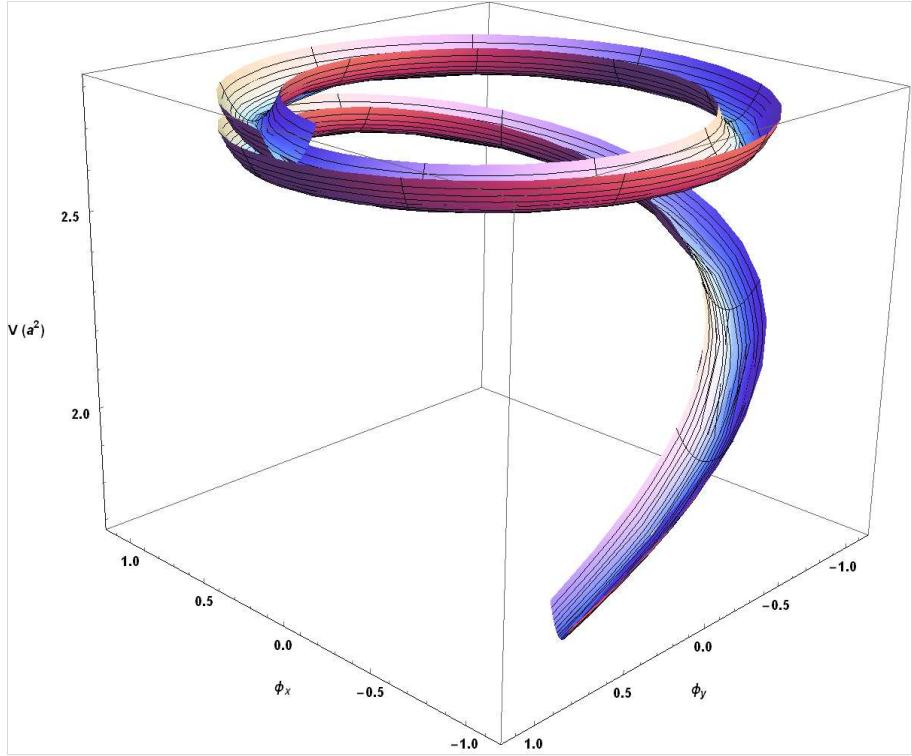}
\caption{The helicoid potential with unit $a^2$ and $c=0.6$. The radial direction has a minimum
at $|\Phi|=1$ where the field norm is strongly stabilized during inflation, while the phase direction is
sufficiently flat to generate slow-roll inflation.
Super-Planckian inflaton excursion is manifest along the phase direction rather than by a physical field.}
\end{figure}

\begin{figure}
\centering
\includegraphics[width=120mm, height=100mm,angle=0]{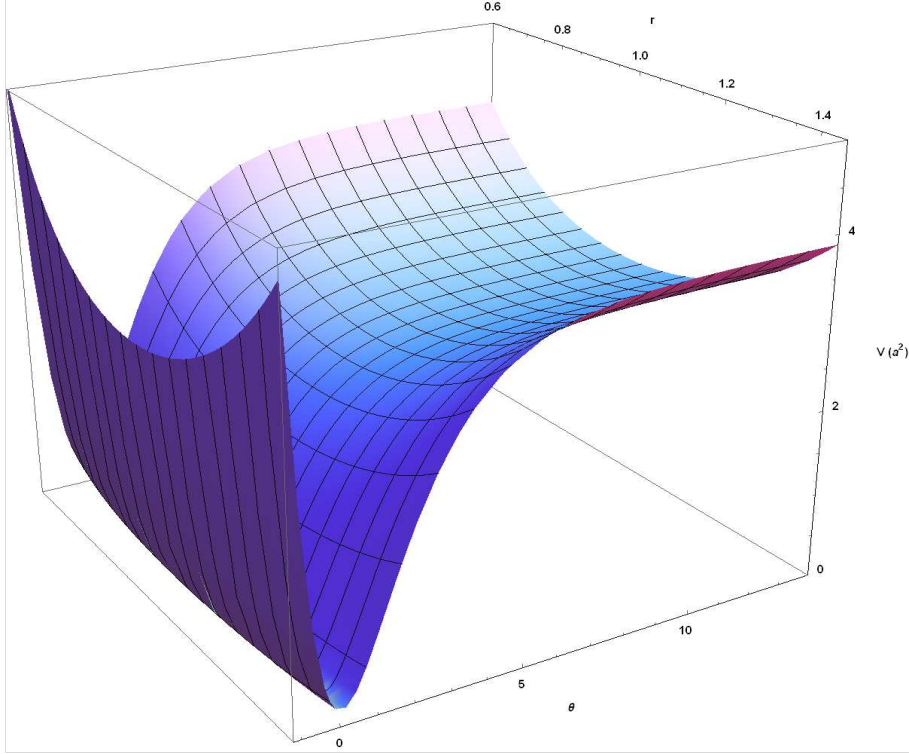}
\caption{The same potential as in Fig.~1 in terms of $r\equiv|\Phi|$ and $\theta\equiv\arg(\Phi)$. The local minimum along $r=1$ is clearly shown in the figure.}
\end{figure}

In $N=1$ supergravity, the F-term scalar potential is determined by the \kah potential $K$ and
the superpotential $W$
\be
V=e^K(K^{i\bar{j}}D_iW D_{\bar{j}}\bar{W}-3W\bar{W}),
\ee
where $K_{i\bar{j}}=\partial_i\partial_{\bar{j}}K$ and $D_iW=\partial_iW+K_iW$.
During inflation the field $X$ is strongly fixed at its vacuum expectation value $\langle X\rangle=0$.
The F-term scalar potential then simplifies to
\be
\begin{split}
V(r,\te)&=e^KD_X W D_{\xb}\wb
=a^2\frac{e^{r^2}}{r^2}(e^{-2c\te}+1-(r^{ic}+r^{-ic})e^{-c\te}) \\
&=a^2\frac{e^{r^2}}{r^2}(e^{-2c\te}+1-2\cos(c\log r) e^{-c\te})
\end{split}
\ee
in Planck units ($M_P=1$), with $\Phi\equiv re^{i\te}$.
The $r-$dependent terms in $V$, $e^{r^2}/r^2$ and $-2\cos(c\log r) e^{-c\te}$ all have minima
at $r=1$, irrespective of $\te$. Therefore, the field norm $|\Phi|$ is strongly stabilized at
$\langle|\Phi|\rangle=1$, and the residual phase-dependent potential reduces to
\be
V(\te)=a^2(1-e^{-c\te})^2, \label{staro}
\ee
with the rescaled parameterization $a\rightarrow a\sqrt{e}$.
The potential $V(r,\te)$ is given in Fig.~1, for the parameter selection $c=0.6$.
The potential shows a deformed helicoid structure. Fig.~2 gives the helicoid potential in terms of $r$ and $\theta$
which clearly shows the potential reaches its local minimum at $r=1$.
In Fig.~1 the minimum in the radial direction represents a deformed helical trajectory along which the inflaton evolves.
Comparing against the helical inflation trajectory for quadratic inflation \cite{Li:2014vpa}, the deformed
path becomes rather steep, finally forcing departure from the inflationary phase, for small $\te$,
while it tends toward extreme flatness for large $\te$.
Taking $c=c_S\equiv\frac{2}{\sqrt{3}}$, after a canonical field rescaling
$\te\rightarrow\frac{1}{\sqrt{2}}\te$, the potential (\ref{staro}) exactly reproduces the Starobinsky
model.  However, there is no implied constraint on $c$, and we do not see a special interpretation
of the value $c_S$ at this stage. The model described by Eqs.~(\ref{set1}) therefore represents
a generalized Starobinsky-like model of inflation. For small
$c\rightarrow0$, it approaches quadratic inflation.  More details on the inflationary predictions of this
potential are given in Fig.~3. These predictions are very well consistent with new
experiment data \cite{Planck:2015xua, Ade:2015oja, Ade:2015tva}, as long as the parameter $c$ is not too small.

\begin{figure}
\centering
\includegraphics[width=100mm, height=100mm,angle=0]{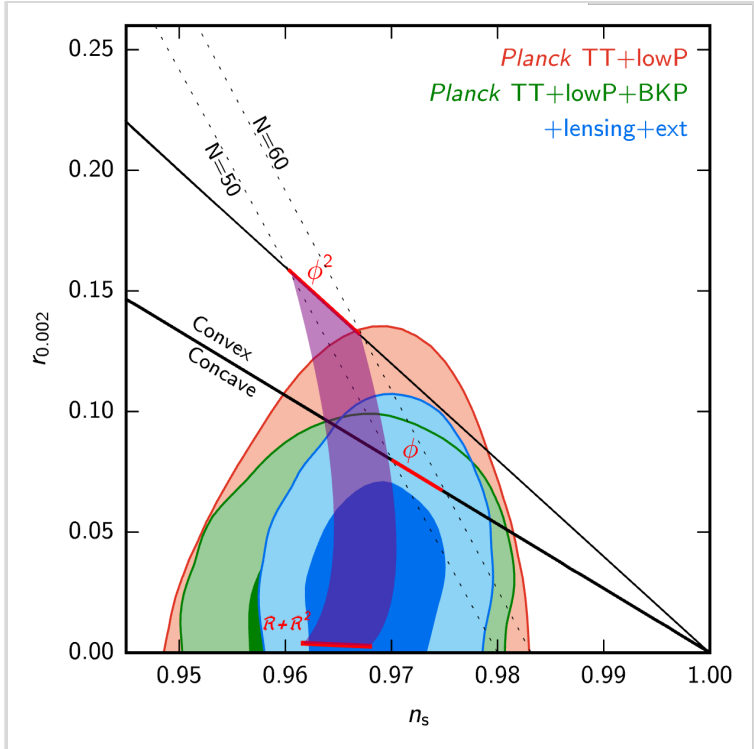}
\caption{The predictions of Starobinsky-like inflation (purple strip), and new experimental data from Ref.~\cite{Planck:2015xua}.}\label{fig3}
\end{figure}

\subsection*{Interpolation from Natural Inflation to Starobinsky-like Inflation}
We have shown that a simple supergravity construction (\ref{set1}) can lead to natural inflation if
the power of $\Phi$ in $W$ is real, or Starobinsky-like inflation if the power of $\Phi$ is pure imaginary. A
natural consideration is the prospect of generalizing the power to arbitrary complex values, as in the
following superpotential
\be
W=a\frac{X}{\Phi}(\Phi^{\chi}-1), \label{set2}
\ee
where $\chi=b+ic$.
It is easy to show that after field stabilization $X\rightarrow\langle X\rangle=0$, the prior
superpotential leads to the scalar potential
\be
V(r,\te)=a^2\frac{e^{r^2}}{r^2}(r^{2b}e^{-2c\te}-2r^b\cos(c\log r+b\te) e^{-c\te}+1).
\ee
Taking a small real exponent, with $c=0$ and $b\ll1$, this potential stabilizes the field norm $|\Phi|\approx1$ and the
inflaton potential reduces to $V(\te)=2a^2(1-\cos (b\te))$, corresponding to
natural inflation, as detailed in \cite{Li:2014unh}.
Curvature along the radial direction is determined by the coefficient $e^{r^2}/r^2$, which admits
a global minimum at $r=1$ and gives a large mass above Hubble scale. Extra couplings between the field norm $r$
and phase $\te$ in $V(r,\te)$ can partially affect the stabilization of $|\Phi|$, although
$b\ll1$ and $e^{-c\te}\ll1$ during inflation, such that corrections to observables are of order $o(b^2)$,
and can be ignored in a primary evaluation. With stabilized field norm $|\Phi|=1$, but no constraint on $c$,
the scalar potential $V(r,\te)$ becomes
\be
V(\te)=a^2(e^{-2c\te}-2\cos(b\te) e^{-c\te}+1).
\ee
By varying $b$ and $c$ one may cleanly interpolate from natural inflation ($c=0, b\ll1$) to Starobinsky-like
inflation ($c>0, b=0$).  Parameterizations of the potential corresponding to the quadratic inflation, natural inflation,
interpolation inflation ($bc\neq0$), and Starobinsky-like inflation scenarios are shown in Fig.~4. The
deformed potentials tend to be flatter in the large field region, and steeper in small field region,
indicating weaker tensor fluctuation.

\begin{figure}
\centering
\includegraphics[width=160mm, height=160mm,angle=0]{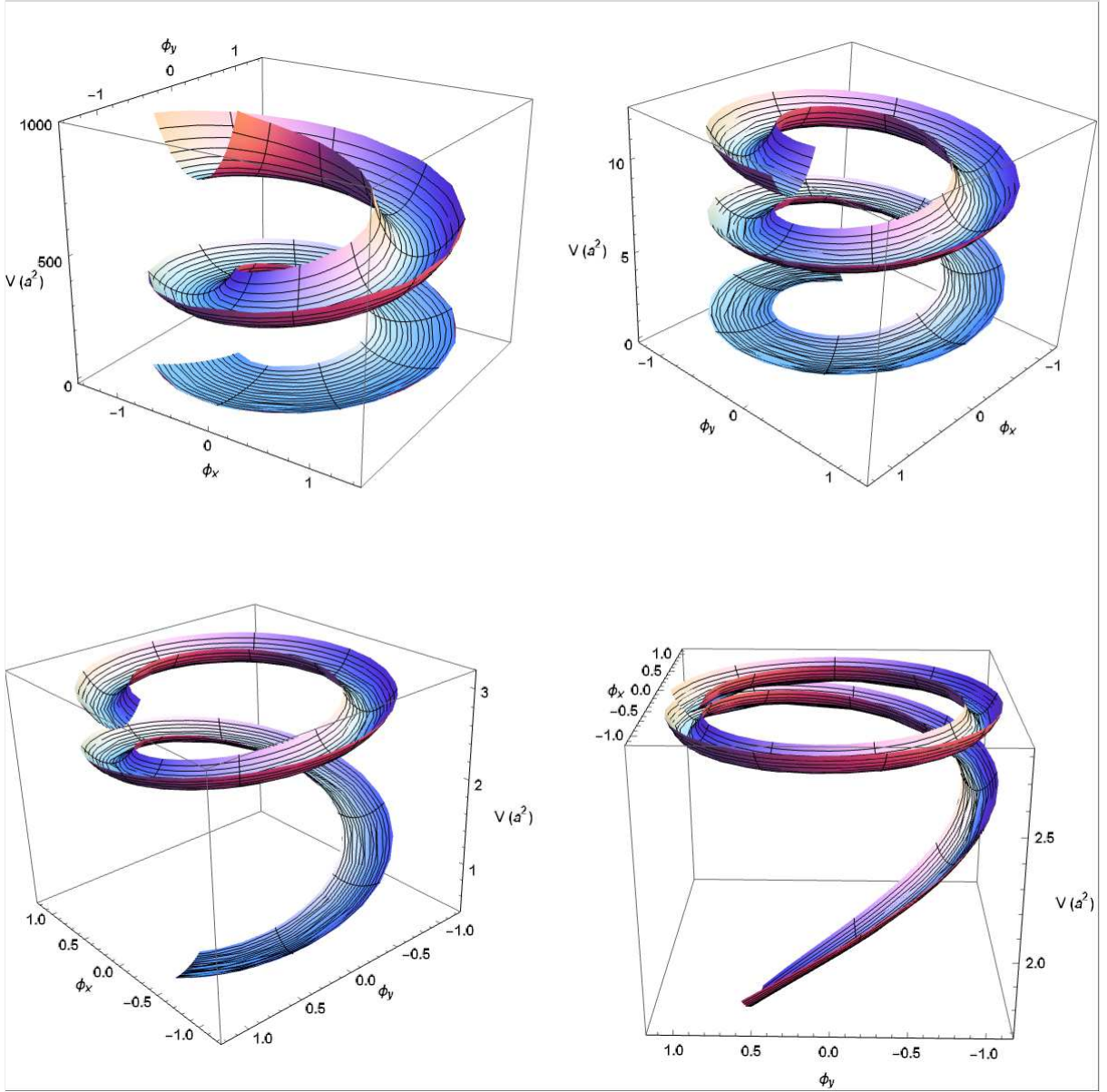}
\caption{Parameterization of the helicoid potentials associated with quadratic inflation (upper-left), natural inflation (upper-right),
interpolation inflation (lower-left), and Starobinsky-like inflation (lower-right) scenarios are depicted. For the later three
inflationary models, the parameters $(c, b)$ are selected as $(0, 0.15)$, $(0.2, 0.15)$, and $(0.6, 0)$, respectively. Deformations render
the helicoid flatter in the large field region, and steeper in small field region.}\label{fig4}
\end{figure}

The potential's deformation can be effectively characterized by the complex parameter $\chi\equiv b+ic$.
Its real component $\Re(\chi)$ relates to the phase/axion decay constant for natural inflation and its
imaginary component $\Im(\chi)$ describes the interpolation between quadratic inflation and
Starobinsky inflation, which is, according to the no-scale supergravity realization of Starobinsky-like inflation \cite{nsp, Ferrara:2013rsa},
the parameter $n$ in the generalized no-scale type \kah potential
\be
K=-n\log(\Phi+\pb)+f(\Phi)+\bar{f}(\pb)
\ee
 that describes a \kah manifold with curvature $R=\frac{2}{n}$. The parameter $\Im(\chi)$
is introduced in Ref.~\cite{Ellis:2013nxa} as a phenomenological generalization of Starobinsky inflation, and it is also
the parameter $\alpha$ in language of
$\alpha$-attractors \cite{Kallosh:2013yoa}.
The ratio $\Re(\chi)/\Im(\chi)$ indicates whether the deformed potential for interpolation inflation better
approximates either natural inflation or Starobinsky-like inflation.
Inflationary predictions of the model (\ref{set2}) with different index values $\chi$ are presented in Fig.~5. As
shown in the graph, with fixed e-folding number $N=60$, each
point on the $n_s-r$ plane \cite{Planck:2015xua} within the region favored by new Planck results
is coincident with predictions of the generalized inflationary model
(\ref{set2}) for some specific choice of the index $\chi$.

\begin{figure}
\centering
\includegraphics[width=100mm, height=100mm,angle=0]{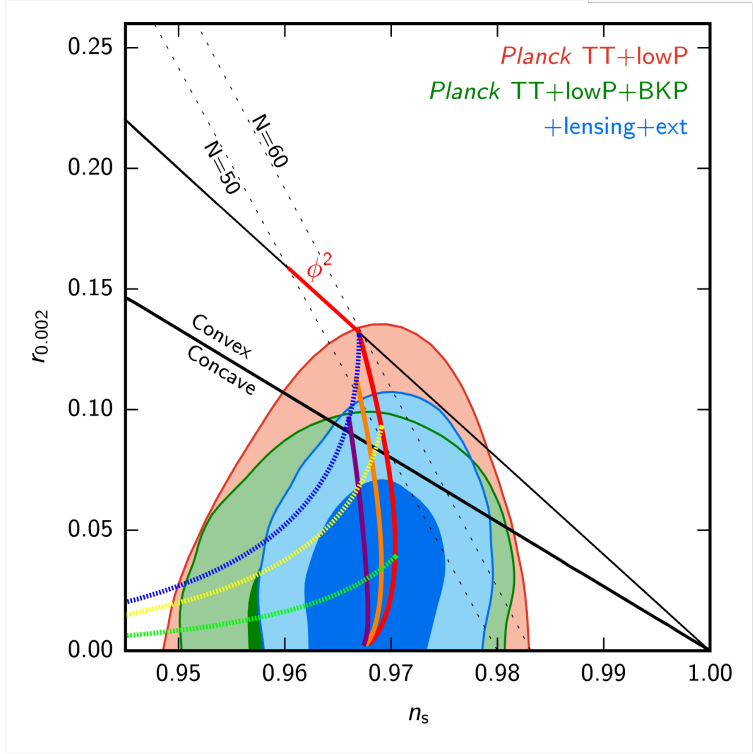}
\caption{Predictions on $n_s-r$ relation of the generalized inflationary model (\ref{set2}) with fixed e-folding number $N=60$.
Dashed lines (with running $b$) represent blue: $c=0$ (natural inflation), yellow: $c=0.05$, and green: $c=0.2$;
thick lines (with running $c$) represent red: $b=0$ (Starobinsky-like inflation), orange: $b=0.1$, and purple: $b=0.14$.}\label{fig5}
\end{figure}

It is surprising that the simple supergravity model (\ref{set2}) can introduce such an abundant variety of results,
which effectively characterize the entire $n_s-r$ region consistent with new Planck data by modulation of the
single complex-valued index $\chi$. However, the origin of this index is a puzzle, and its physical meaning is
unclear from the model (\ref{set2}).  It should also be noted that this model exhibits a pole at $\Phi=0$, and is
well-defined only in the large field region $|\Phi|\gg0$; as such, it should be considered as an effective
theory with heavy fields integrated out.  A more fundamental model including heavy fields may help us to
unveil the physical meaning of the index $\chi$.

\section{Global $U(1)$ Symmetry in the Superpotential}
The supergravity construction for helical phase inflation is considered to be an effective theory.
The \kah potential and superpotential are rather simple, while several critical facets of the
superpotential, such as the pole at $\Phi=0$ and the phase monodromy, require further elaboration.
We follow the method proposed in \cite{Li:2014vpa, Li:2014unh}, where the phase monodromy appearing in
the superpotential (\ref{set1}) and (\ref{set2}) is realized by explicitly breaking of a global $U(1)$
symmetry.

We begin with the following superpotential
\be
W=aX\Psi(e^{-\alpha T_1}-\rho)+Y(e^{-\beta T_2}-\sigma\Psi)+Z(\Phi\Psi-\lambda)+\cdots,
\label{set3}
\ee
in which the first term $W_I=aX\Psi(e^{-\alpha T_1}-\rho)$ is used to generate the inflaton potential, so the
coefficient should be rather small $a\ll1$. Another pair of terms appear at energy scales hierarchically higher than
that of inflation, with coupling coefficients significantly larger than $a$, which we presently set to $1$ for convenience.
Two racetrack-type non-perturbative terms are included.
Given $T_1=T_2$, the superpotential (\ref{set3}) reduces to the natural inflation model
\cite{Li:2014unh}.
The exponentials are multiplied with stabilizer fields $X$, $Y$ that vanish during inflation and their
F-terms provide for the inflaton potential or for field stabilization.  The non-perturbative terms are expected to
be obtained from D-brane instanton effects. The D-brane instanton effects are widely studied in
the construction of matter couplings favored for their phenomenological aspects (more details are provided in
\cite{Ibanez:2012zz}). The advantage of the D-brane instanton mechanism is that the magnitudes of its associated
terms do not have to be too small and can be applied for field stabilization above the inflation scale. In contrast, the
non-perturbative superpotential from gauge theory instantons is usually substantially suppressed and
there is not much space to establish a hierarchy between inflation and field stabilization.
Extra \kah moduli terms are omitted in (\ref{set3}), which are expected to provide linear
constraints on $T_1$ and $T_2$. A detailed study of these terms will be provided later.

The superpotential (\ref{set3}) features a global $U(1)$ symmetry.
The superfields transform under the $U(1)$ symmetry
\be
\begin{split}
&X\rightarrow Xe^{-iq\te}, ~~Y\rightarrow Ye^{-iq\te}, ~~Z\rightarrow Z, \\
&\Psi\rightarrow \Psi e^{iq\te}, ~~~~\,\Phi\rightarrow\Phi e^{-iq\te}, ~~\,T_2\rightarrow T_2-i\frac{q}{\beta}\te,
\end{split}
\ee
Given $T_1=T_2$, the superpotential (\ref{set3}) reduces to the natural inflation model
\cite{Li:2014unh}. The $U(1)$ symmetry is explicitly broken by the inflation term $W_I$, while
we can assume in this model that $T_1$ is neutral under $U(1)$.  A global $U(1)$ symmetry of this kind
appears in the \kah potential more naturally. For the matter fields, their \kah potentials are of the minimal type, and invariant
under $U(1)$ transformation. The \kah modulus $T_2$ shifts under $U(1)$, and its \kah potential is of
no-scale type, which is thus independent of the imaginary component. Consequently, the global $U(1)$ symmetry
is inherited by the F-term scalar potential, forming an exactly flat direction. To lift the flat direction
one has to break the global $U(1)$ symmetry, which can be achieved by applying a linear constraint between
two \kah moduli
\be
T_1-\kappa T_2+\delta=0 \,. \label{cstr}
\ee
Here, we require the constraint to satisfying the conditions $\Re(T_1)>0$ and $\Re(T_2)>0$, since the real
components of $T_i$ give the volumes of internal cycles. We will show that such linear constraint on \kah
moduli are common in non-geometric flux compactification, and that the coefficient $\kappa$ has a clear geometric origin.

The flat direction is lifted after $U(1)$ symmetry breaking, leaving a unique Minkowski vacuum. To
isolate the vacuum, we need to solve the equations:
\be
W=D_zW=W_z+K_zW=W_z=0, \label{mink}
\ee
where $z\in\{\Phi, \Psi, X, Y, Z, T_1, T_2\}$. It is easy to show that the superpotential (\ref{set3})
admits a supersymmetric Minkowski vacuum at
\be
\begin{split}
&X=Y=Z=0, ~~ T_1=-\frac{\ln\rho}{\alpha}, ~~T_2=\langle T_2\rangle=-\frac{\ln\rho}{\alpha\kappa}+\frac{\delta}{\kappa}, \\
&\Psi=\frac{1}{\sigma}e^{-\beta\langle T_2\rangle}, ~~~~~\,\Phi=\sigma\lambda e^{\beta \langle T_2\rangle}.
\end{split}
\ee
The parameters are manually adjusted such that $\langle|\Phi|\rangle\gg\langle|\Psi|\rangle$.
The fields $Y, Z, \Psi, T_2$ obtain masses significantly above the inflationary energy scale
according to the superpotential (\ref{set3}), whereas $T_1$ is limited by the prior constraint.
During inflation these degrees of freedoms are thus frozen and should be integrated out.
To integrate out heavy fields, we should solve F-term equations.
Non-trivial results are obtained from the F-term equations associated with the stabilizer fields $Y$ and $Z$:
\be
\begin{split}
&F_Y=D_YW=e^{-\beta T_2}-\sigma\Psi+\bar{Y}W\approx e^{-\beta T_2}-\sigma\Psi=0, \\
&F_Z=D_ZW=\Phi\Psi-\lambda+ \bar{Z}W\approx \Phi\Psi-\lambda=0,
\end{split}
\ee
where we have ignored terms proportional to $\bar{Y}$ or $\bar{Z}$, given that these fields
are strongly stabilized
at $\langle Y\rangle=\langle Z\rangle=0$ during inflation.
The heavy fields are solved for in term of $\Phi$
\be
\begin{split}
\Psi&=\frac{\lambda}{\Phi}, \\
e^{-\beta T_2}&=\frac{\sigma\lambda}{\Phi}, \\
e^{-\alpha
T_1}&=e^{\alpha\delta}(\sigma\lambda)^{\frac{\alpha\kappa}{\beta}}\Phi^{-\frac{\alpha\kappa}{\beta}},
\end{split}
\ee
and the effective low energy theory becomes
\be
W=a\lambda
\frac{X}{\Phi}(e^{\alpha\delta}(\sigma\lambda)^{\frac{\alpha\kappa}{\beta}}\Phi^{-\frac{\alpha\kappa}{\beta}}-\rho),
\ee
which is the superpotential (\ref{set2}) for helical phase inflation, with a suitable parameter
redefinition. The index $\chi$ is given by $\chi=-\frac{\alpha}{\beta}\kappa$,
where $\alpha$ and $\beta$ are positive parameters and only affect the magnitude of $\chi$.

\section{Moduli Constraint from flux compactification}

Constraints on \kah moduli in type \MyRoman{2}B compactification, or on the T-dual complex structure moduli
in type \MyRoman{2}A compactification,
are obtained from moduli stabilization, which requires \kah moduli couplings in the superpotential.
However, no perturbative term on \kah moduli can be generated from type \MyRoman{2}B compactification
with RR or NSNS fluxes. By contrast, they do appear in type \MyRoman{2}A compactifications with geometric
fluxes and also in type \MyRoman{2}B compactifications with non-geometric fluxes. Both geometric and
non-geometric fluxes arise from T-duality of NSNS $3$-form fluxes.

\subsection{Type \MyRoman{2}A Compactification with Geometric Fluxes}
T-duality connects \MyRoman{2}B and \MyRoman{2}A orientifold compactifications.
The linear constraint on \kah moduli in \MyRoman{2}B compactification is T-dualized to a linear
constraint on complex structure moduli in \MyRoman{2}A.
For type \MyRoman{2}B orientifold compactification with NSNS flux $H_{ijk}$, taking T-duality along
a direction $x_i$ of the internal space $M$, the flux $H_{ijk}$ is mapped into a new type of flux.
This is the geometric flux $\omega^{i}_{jk}$, which is equivalent to \MyRoman{2}A compactification on a twisted torus.
The twisted torus is described by the geometric flux $\omega^{i}_{jk}$ as
\be
d\eta^i=-\frac{1}{2}\omega^{i}_{jk}\eta^j\wedge\eta^k,
\ee
where $\omega^{i}_{jk}=-\omega^{i}_{kj}$ and $\eta^j$ are tangent 1-forms linearly depending on
the internal coordinates $x^j$. The dual tangent vectors $Z_i$ of the 1-form $\eta^i$ form a Lie algebra with
geometric fluxes $\omega^{i}_{jk}$ as the structure constants
\be
[Z_i, Z_j]=\omega_{ij}^kZ_k \,. \label{lieA}
\ee
Consequently, the fluxes should satisfy the Jacobi identity
\be
\omega^{i}_{[jk}\omega^{l}_{m]n}=0 \,. \label{const1}
\ee

For \MyRoman{2}A toroidal compactification with O6-planes and geometric fluxes $\omega$, the
orientifold action is $\Omega (-1)^{F_L}R_A$ and the involution $R_A$ acts on the local internal complex
coordinates $z_i$ as $R_A(z_i)=\bar{z_i}$.
In the low-energy effective theory, the NSNS and geometric fluxes generate following terms in
superpotential
\be
W_{NS\&\omega}=\int_{T_6}\Omega_c\wedge(H_3+dJ_c),
\ee
where $\Omega_c$ and $J_c$ are a holomorphic 3-form and \kah 2-form, respectively. Also, the RR flux
superpotential is of the typical form
\be
W_{RR}=\int_{T_6}e^{J_c}\wedge F_{RR},
\ee
where the $F_{RR}$ are RR fluxes. Combining $W_{NS\&\omega}$ and $W_{RR}$, the full superpotential reads
\be
W(T_i, S, U_i)=P_{-1}(T_i)+SP_0(T_i)+\sum_{k=1}^{3}U_kP_k(T_i),
\ee
where $P_{-1}$ is cubic on $T_i$, while $P_0, P_i$ are linear on $T_i$. Geometric flux quanta as
coefficients in $W$ are subjected to the Jacobi identity constraint (\ref{const1}). The tree level
\kah potential is of no-scale type
\be
K=-\log(S+\bar{S})-\sum_{i=1}^{3}\log(U_i+\bar{U_i})-\sum_{i=1}^{3}\log(T_i+\bar{T_i}).
\ee

We are interested in obtaining a supersymmetric Minkowski vacuum from $W$. Supersymmetric Minkowski
vacuum equations (\ref{const1}) for $S$ and $U_i$,
together with the condition $W=0$,
require
\be
P_{-1}(T_i)=P_0(T_i)=P_i(T_i)=0.
\ee
In principle, the three \kah moduli can be solved for from the prior equations. However, there are more equations
than variables, so mutual consistency is not guaranteed.  On the other hand, we do
have substantial freedom to turn on fluxes, and the two extra equations are equivalent to non-linear constraints
on these flux coefficients. Since the fluxes are quantized, only integer solutions of these constraints are
physical. Additionally, there are three equations associated with the supersymmetric Minkowski vacuum for $T_i$
\be
\partial_{\,T_i}P_{-1}+S\partial_{\,T_i}P_{0}+\sum_{k=1}^3 U_k\partial_{\,T_i}P_{k}=0 \label{const2}
\ee
with $4$ variables $S$ and $U_k$. Through variable elimination, one obtains a constraint relating two
moduli, of the type needed in Eq.~(\ref{cstr}).

However, both $P_0$ and $P_i$ are linear on $T_i$, and their coefficients are from flux quanta, and thus real.
The expressions in (\ref{const2}) give a moduli constraint corresponding to a real parameter $\kappa$ in Eq.~(\ref{cstr}),
yielding a real value for the index $\chi$, which produces natural inflation. This is equivalent to the simple isotropy condition
stating that the three sub-tori of $T^6=T^2\times T^2\times T^2$ admit exchange symmetry with $T_i=T$, and $U_i=U$. To
obtain more general moduli constraints with complex coefficients and a generically complex $\kappa$, we need
$P_0$ or $P_i$ to exhibit high-order couplings. This is realized by introducing non-geometric
Q-fluxes, which are T-duals of the geometric flux $\omega$.

\subsection{Type \MyRoman{2}B Compactification with Q-fluxes}
By taking successive T-dualities along different compact coordinates, more fluxes arise without
geometric interpretation, which are dubbed non-geometric fluxes. These fluxes are related to each
other under T-duality
\be
H_{ijk}\stackrel{T_i}{\longleftrightarrow}\omega_{jk}^i\stackrel{T_j}{\longleftrightarrow}Q_{k}^{ij}\stackrel{T_k}{\longleftrightarrow}R^{ijk},
\ee
where $T_i$ refers to the T-duality along compact direction $x^i$. In this section,
we study type \MyRoman{2}B orientifold compactification on Calabi-Yau manifolds $M$ with
an orientifold projection $\Omega_P(-1)^{F_L}R_B$, in which $\Omega_P$ is the world-sheet parity
operator, $F_L$ is the left-moving fermion number and $R_B$ is the orientifold involution.
The superpotential from non-geometric fluxes contains \kah moduli couplings at the perturbative level and
admits rich vacuum configurations.

For type \MyRoman{2}B orientifold compactification with O3-planes,
the orientifold involution $R_B$ acts on compact coordinates $x_i$ as $R_B: x_i\rightarrow -x_i$,
and its actions on
the \kah form $J$ and the holomorphic 3-form $\Omega_3$ are therefore
\be
R_B(J)=J, ~~~~~~~~~ R_B(\Omega_3)=-\Omega_3.
\ee
The involution $R_B$ projects out even parts of the cohomology $H^{3}(M)$ and odd parts of the cohomologies
$H^{1,1}(M)$ and $H^{2,2}(M)$. Remaining cohomologies are denoted as
\be
\begin{split}
{\omega_i}&\in H_+^{1,1}(M)                         ~~~~~~~~~~~~i=1,\cdots, h_+^{1,1}  \\
{\tilde{\omega}^i}&\in H_+^{2,2}(M)                 ~~~~~~~~~~~~i=1,\cdots, h_+^{1,1}  \\
\{\alpha_m, \beta^m\}&\in H_-^3(M)  ~~~~~~~~~~~~~m=0, \cdots, h_-^{2,1}.
\end{split}
\ee
The holomorphic 3-form $\Omega_3$ can be expanded in terms of the symplectic basis of $H_-^3(M)$
\be
\Omega_3=X^m\alpha_m-F_m\beta^{m},
\ee
where $X^m=\int_{A^m}\Omega_3$ and $F_m=\int_{B_m}\Omega_3$ are periods of the compactification
manifold $M$ with a symplectic three cycle basis $\{A^m,~B_m\}$.
Taking $X^m$ as coordinates on the complex structure moduli space, periods $F_m$ can be represented as partial
derivatives on the prepotential $F=f_{ijk}X^iX^jX^k/X^0$, specifically $F_m=\partial_{X_m} F$.
The low-energy effective theory is described by $N=1$ supergravity with a tree level \kah potential
\be
K=-\log(-i\int_M \Omega_3\wedge\bar{\Omega_3})-\log(S+\bar{S})-2\log(e^{-\frac{3}{2}\phi}\int_M J\wedge
J\wedge J).
\ee
The superpotential is given by the generalized Gukov-Vafa-Witten superpotential \cite{Blumenhagen:2015kja}
\be
W=\int_M(F_3-iSH+iT_i(Q\bullet\tilde{\omega}^i)+\cdots)\wedge\Omega_3, \label{sup1}
\ee
where the $Q$ action on the $p$-form $F_{M_1\cdots M_p}$ gives the $p-1$-form
\be
(Q\bullet F)_{NM_1\cdots M_{p-2}}=\frac{1}{2}Q^{JK}_{[N} F_{M_1\cdots M_{p-2}]JK},
\ee
leading to Q-flux terms in the superpotential that depend linearly on the \kah moduli $T_i$.
The 4-forms $\tilde{\omega}^i$ are a basis of even (2, 2)-cohomology.
The first two terms in (\ref{sup1}) are just the well-known Gukov-Vafa-Witten superpotential
\cite{Gukov:1999ya} for NSNS and RR fluxes, namely
$F_3=dC_2$ and $H=dB_2$ .
In the superpotential $W$ we have ignored the geometric and R-type fluxes that could also play
an interesting role in moduli stabilization.  The R-type fluxes do not appear in the superpotential due
to the symmetry under orientifold projection but they can involve in the D-terms potential
\cite{Robbins:2007yv, Shukla:2015rua, Shukla:2015bca}.
More details on these fluxes are provided in Ref.~\cite{Blumenhagen:2015kja}.

Expanding the $p$-form fluxes on a cohomology basis with arbitrary flux quanta, the superpotential $W$ can
be expressed in terms of flux quanta and moduli
\be
W=-(e_mX^m-\tilde{e}^mF_m)+iS(a_mX^m-\tilde{a}^mF_m)+iT_i(b_{mi}X^m-\tilde{b}^{mi}F_m)+\cdots.
\label{sup2}
\ee
The superpotential $W$ implicitly depends on the complex structure moduli $U^m$ through the
$H_-^{2,1}(M)$ periods $X^m$:
\be
U^m=-i\frac{X^m}{X^0}.
\ee

After turning on non-geometric fluxes, the Lie algebra (\ref{lieA}) is extended with new generators. The
NSNS, geometric and non-geometric fluxes become structure constants of the extended Lie algebra, and the
Jacobi identities of the Lie algebra introduce new constraints on fluxes \cite{Shelton:2005cf, Aldazabal:2006up, Blumenhagen:2015kja}.
Additionally, these fluxes also
contribute to the RR $4$-form and $8$-form tadpoles, which should satisfy the tadpole cancellation
conditions in conjunction with the O3/D3 and O7/D7 contributions.

The superpotential in Eq.~(\ref{sup2}) contains perturbative couplings of the \kah moduli. In contrast with the
case of type \MyRoman{2}A orientifold compactification with geometric fluxes, the complex structure
terms $b_{mi}X^m-\tilde{b}^{mi}F_m$ coupled with \kah moduli are not just linear, but are up to third order,
leading to more interesting moduli stabilization and vacuum configurations. The supersymmetric
Minkowski vacuum and moduli stabilization from the Type \MyRoman{2}B compactifications on
orientifolds have been studied in Refs.~\cite{Aldazabal:2006up, Shelton:2006fd, Micu:2007rd,
Font:2008vd, Guarino:2008ik, deCarlos:2009qm, Aldazabal:2011yz}. Most of these works targeted
the compactification on isotropic $T^6$, i.e., with an exchange symmetry among three sub-tori so that
$T_1=T_2=T_3$ and $U_1=U_2=U_3$. In our study, the multi-\kah and complex structure moduli are needed.
We want to realize constraints between \kah moduli instead of fixing all the \kah moduli completely. Without
the isotropy constraint, there is additional freedom to adjust the flux quanta, producing richer vacuum
configurations.
However, with more \kah and complex structure moduli, the Jacobi identities for fluxes become extremely
clumsy. The Jacobi identities, together with the tadpole cancellation conditions and equations for
the supersymmetric Minkowski vacuum, could be solved by numerically scanning the parameter space. Mathematical
techniques from algebraic geometry have been applied to solve the flux constraints and
supergravity equations using the programs {\it Mathematica} and {\it Singular}
\cite{Gray:2008zs}. In this work, we propose a toy model with a superpotential like that in Eq.~(\ref{sup2}) to show
how the constraint on \kah moduli appears through moduli stabilization in the supersymmetric Minkowski
vacuum.  However, in this toy model we do not expect to solve the Jacobi identities with given fluxes.
We will discuss the effects of these Jacobi identities on the Minkowski vacua later.

We consider the Type \MyRoman{2}B orientifold compactification on a Calabi-Yau manifold with
$h_+^{1,1}=2, ~h_-^{1,1}=0, ~h_-^{1,2}=3$. In this model, the \kah potential is
\be
K=-\sum_{i=1}^3\log(U_i+\bar{U}_i)-\log(S+\bar{S})-\log(T_1+\bar{T}_1)-2\log(T_2+\bar{T}_2),
\ee
where we have implicitly assumed that in the prepotential $F$
the only non-vanishing component of
symmetric coefficients $f_{ijk}$  is $f_{123}=1$.
The fluxes are adjusted to generate following superpotential
\be
\begin{split}
W&=W_{NSR}+W_Q, \\
W_{NSR}&=e_0+q_1U_2U_3+a_1SU_1+a_2(U_1+S)(U_2+U_3)+\tilde{h}SU_1U_2U_3, \\
W_Q&=(U_2-U_3)(bT_1-i\tilde{b}T_2U_1). \label{sup3}
\end{split}
\ee
The superpotential $W_{NSR}$ admits exchange symmetries
in terms of $U_2\leftrightarrow U_3$ and $S\leftrightarrow U_1$. Only even order couplings among
$S$ and $U_i$ are considered, so that the flux quanta appear in $W_{NSR}$ as real coefficients. These
limitations on the superpotential are not necessary for a supersymmetric Minkowski vacuum, but make the
calculations sufficient simple for an example. Another term $W_Q$ is quadratic in $U_i$ and is expected to
be obtained from $Q$-fluxes, which can generate
couplings of the complex structure moduli up to third order, as shown in Eq.~(\ref{sup2}).
Again we remind the readers that above potentials should be considered as an example to show the linear constraint of K\"aher moduli obtained as relic of moduli stabilization, instead of generating a ``physical" Minkowski vacuum since not all Jacobi identities are fully satisfied with given fluxes and prepotential coefficient.

It is straightforward to show that the superpotential $W_{NSR}$ yields a supersymmetric Minkowski vacuum
\cite{Kachru:2002he}, i.e. that
\be
W_{NSR}=\partial_{U_i}W_{NSR}=\partial_S W_{NSR}=0, \label{smk}
\ee
at
\be
\begin{split}
S&=U_1=\sqrt{\frac{q_1}{a_1\tilde{h}}}\Big(-a_1\pm2a_2\sqrt{\frac{a_1}{q_1}}\Big)^{1/2}, \\
U_2&=U_3=\pm\sqrt{\frac{a_1}{q_1}}S~,~
e_0~=~\frac{q_1}{a_1\tilde{h}}\Big(-a_1\pm2a_2\sqrt{\frac{a_1}{q_1}}\Big)^{2}~.~
\end{split}
\ee
Taking the flux quanta $(e_0, a_1, a_2, q_1, \tilde{h})=(2,~2,~-2,~2,~2)$, and ignoring unphysical
solutions,
one can realize a supersymmetric vacuum at $S=U_i=1$. Examples with complex vacuum expectation values of $U_i$
are also provided in \cite{Kachru:2002he}.

If the $Q$-fluxes introduce perturbative couplings $W_Q$ of \kah moduli, then the equations for
a supersymmetric Minkowski vacuum become
\be
\begin{split}
\partial_S W_{NSR}&= W_{NSR}+(U_2-U_3)(bT_1-i\tilde{b}T_2U_1)=0, \\
\partial_{T_1}W_Q&=\partial_{T_2}W_Q=U_2-U_3=0,\\
\partial_{U_1}W_{NSR}-i\tilde{b}T_2(U_2-U_3)&=0, \\
\partial_{U_2}W_{NSR}+(bT_1-i\tilde{b}T_2U_1)&=0, \\
\partial_{U_3}W_{NSR}-(bT_1-i\tilde{b}T_2U_1)&=0.
\end{split}
\ee
It is easy to see that above equations are equivalent to the equations of $W_{NSR}$ (\ref{smk}) plus an
extra constraint on the \kah moduli
\be
bT_1-i\tilde{b}T_2\langle U_1\rangle=0.
\ee
According to the moduli stabilization from $W_{NSR}$, namely $\langle U_1\rangle=1$, this constraint
enforces an imaginary ratio $\kappa=i\tilde{b}\langle U_1\rangle/b$ and yields an imaginary
index $\chi$, which produces Starobinsky-like helical phase inflation. For the models with complex $\langle
U_1\rangle$, we may realize a complex index $\chi$ as well, which produces interpolation inflation. However, the
pure imaginary $\langle U_1\rangle$ for natural inflation corresponds to a boundary of the complex structure
moduli space. This solution is not physical, as it indicates a degenerate internal space.

The overall superpotential contains term associated with both the string moduli (\ref{sup3}) and the
helical phase inflation supergravity construction (\ref{set3}), and the latter also depend
on the \kah moduli through non-perturbative effects.  Therefore, non-perturbative terms appear in equations
$\partial_{T_i}W=0$ as well, and may affect the vacuum equations $\partial_{T_i}W=U_2-U_3=0$ that are
necessary to facilitate the exchange symmetry $U_2\leftrightarrow U_3$ in $W_{NSR}$. Fortunately, this
is avoided due to vanishing of the stabilizer fields $X$ and $Y$ in (\ref{set3}). The string moduli
stabilization and generation of dynamics for helical phase inflation are therefore reducible, even though the
\kah moduli appear in both sets of equations.
It is easy to generalize the $Q$-flux superpotential $W_Q$ to get different \kah moduli constraints.
Discrete symmetries are employed here, and they play an important role in simplifying calculations. Since there
are more equations than variables, these discrete symmetries help to maintain mutual consistency
of the equations.  For a more realistic model, i.e. one combining the vacuum
equations with Jacobi identities and RR $4$-form $C_4$, $8$-form $C_8$ tadpole constraints, the
calculations become quite cumbersome, and it is necessary to scan the parameter space numerically
in order to identify realistic vacua
that satisfy the formalism and provide a linear constraint on \kah moduli. In fact, such a constraint on
the \kah moduli is a natural outcome for the supersymmetric Minkowski vacua.

It is interesting to compare our proposal with the work in \cite{Blumenhagen:2015qda}, which also
studied the Starobinsky-like inflation based on non-geometric flux compactification. In \cite{Blumenhagen:2015qda}
the inflaton is from string moduli which strongly interact with other heavy fields, so it needs
to adjust the parameters carefully so that the inflation direction is sufficient flat while all the extra fields
are at or above Hubble scale. Because for large field inflation a super-Planckian field excursion is required
and the inflation dynamics is rather sensitive to the super-Planckian physics, one also needs to check
whether the large field inflation can be reasonably studied in the low energy effective theory.
In particular the following hierarchy should be unbroken under super-Planckian string moduli/axion field excursion \cite{Blumenhagen:2015qda}
\be
M_P>M_s>M_{KK}>M_{moduli}>H_{inflation}>M_{inflaton}. \label{hie}
\ee
Interestingly, in helical phase inflation, the flatness condition of inflation potential is protected by a mildly broken
$U(1)$ symmetry, and the UV-completion problem of large
field inflation is avoided. Thus, we do not need to struggle with the $\eta$ problem
or the effectiveness of the low energy theory obtained from string compactification. Moreover, besides the normal
conditions on string compactification, the above hierarchy condition (\ref{hie})
is replaced by a single hierarchy in Eq.~(\ref{set3}): $a\ll1$. In other words, the first term in Eq.~(\ref{set3}) is
significantly smaller than the other terms so that we can safely integrate them out.
In contrast, in our proposal the non-geometric fluxes lead to much more complex constraints than those in \cite{Blumenhagen:2015qda}.
We expect the constraints from the flux Jacobi identities and Minkowski vacuum conditions can be solved numerically.

Supersymmetric Minkowski vacua with a constraint on the \kah moduli are expected to appear from type
\MyRoman{2}B orientifold compactification on $T^6/\Omega_P(-1)^{F_L}R_B$.
Such compactifications have been studied in Refs.~\cite{Aldazabal:2006up, Shelton:2006fd, Micu:2007rd,
Font:2008vd, Guarino:2008ik,
deCarlos:2009qm, Aldazabal:2011yz, Hassler:2014mla, Blumenhagen:2015kja} as a mechanism for obtaining
vacua with full moduli stabilization. The authors of these works have mainly focused on the
simplified case where an exchange symmetry exists among three sub-tori. The associated results are very limited
unless a non-geometric $P$-flux arising from S-duality of type \MyRoman{2}B string theory is also introduced.
In our case, with loose moduli stabilization criteria and no exchange
symmetry, there is more freedom to arrange the flux quanta in order to obtain supersymmetric Minkowski vacua.
The superpotential for NSNS, RR and $Q$-type fluxes is
\be
W(U_i, S, T_i)=P_{-1}(U_i)+SP_0(U_i)+\sum_{k=1}^{3}T_kP_k(U_i),
\ee
where the terms $P_0$ and $P_i$ for NSNS 3-form and Q-fluxes are cubic in $U_i$:
\be
\begin{split}
P_{-1}&=f_0+i\sum_{i=1}^3f_iU_i-\sum_{i=1}^3\frac{\tilde{f}_i}{U_i}U_1U_2U_3+i\tilde{f}_0U_1U_2U_3,
\\
P_{0}&=ig_0-\sum_{i=1}^3g_iU_i+i\sum_{i=1}^3\frac{\tilde{g}_i}{U_i}U_1U_2U_3-\tilde{g}_0U_1U_2U_3,
\\
P_{k}&=-ih_{k}-\sum_{i=1}^3h_{ik}U_i+i\sum_{i=1}^3\frac{\tilde{h}_{ik}}{U_i}U_1U_2U_3+\tilde{h}_kU_1U_2U_3.
\end{split}
\ee
The supersymmetric Minkowski vacuum equations $W=\partial_S W=\partial_{T_i}W=0$
require
\be
P_{-1}=P_0=P_i=0.
\ee
As is the case for type \MyRoman{2}A orientifold compactification, these equations give vacuum
expectation values for $U_i$ and also certain non-linear constraints on flux quanta.
There are still three vacuum equations from $U_i$ ($\partial_{U_i} W=0$) with four undetermined moduli
$S$ and $T_i$. After variable elimination, we get a linear constraint on two of the \kah moduli.
Since $P$s are cubic in $U_i$ with both real and imaginary coefficients, generically the ratio of \kah
moduli in the constraint is complex.
The non-geometric flux compactification provides an interesting correlation with helical phase
inflation.  Specifically, for type \MyRoman{2}B orientifold compactification on $T^6/\Omega_P(-1)^{F_L}R_B$,
the supersymmetric Minkowski vacua are consistent with helical phase inflation equipped with a certain
index $\chi$.

Given the relationship between Type \MyRoman{2}B orientifold compactification with
non-geometric fluxes and helical phase inflation, it is interesting to study the distribution of
supersymmetric Minkowski vacua in the moduli space.  Proceeding, we assume
that each type of flux quanta forms a direction of the moduli space.
Flux quanta are integral, and the vacua
are thus located on a lattice of the moduli space.
Possible values of the index $\chi$ are uniquely determined in each vacuum, and are therefore discrete points
distributed on the complex plane.  The magnitude of the index $\chi$ also depends upon non-perturbative effects that
may be computed from details of the D-brane configurations. We have shown that the index $\chi$
characterizes the $n_s-r$ plane with fixed e-folding number $N$. In principle, predictions for the $n_s-r$
relation from helical phase inflation should then likewise correspond to isolated points in the $n_s-r$ plane,
significantly reducing associated uncertainties.  Unfortunately, the preferred e-folding
number window $N\in[50, 60]$ introduces an uncertainty $\Delta N\approx10$,
which disperses predictions for the $n_s-r$ relation across in a small region, even when the index $\chi$ is fixed.

The realization of a distribution of supersymmetric Minkowski vacua from non-geometric flux
compactifications can be considered as a generalization of the systematic study of
the vacua emerging from type \MyRoman{2}B compactification with NSNS and RR fluxes in Ref.~\cite{DeWolfe:2004ns}.
The objectives of that work are determination supersymmetric Minkowski vacua fraction favored by low scale
supersymmetry and isolation of their corresponding discrete symmetries. The authors focus mainly on $T^6$
compactifications with exchange symmetry among sub-tori.  In \cite{DeWolfe:2004ns},
techniques from number theory are employed in order to enumerate the vacua according to the integral nature of flux quanta.
Even though the equations for supersymmetric Minkowski vacua ($W=\partial_{U_i}W=0$ in this case)
provide strict limitations, there are still abundant distinct solutions corresponding to variation of the flux quanta.
In their work, the  \kah moduli do not appear in superpotential.
In our case, non-geometric fluxes are introduced, which are constrained by
Jacobi identities. Equations for the determination of vacua are similar in both cases, although the variety of
solutions is expected to be richer in our model, given the absence of exchange symmetry among sub-tori.
On the other hand, the calculations become dramatically more involved in this case.
In this work, we have shown that beyond the motivations discussed in \cite{DeWolfe:2004ns},
the distribution of supersymmetric Minkowski vacua
is also deeply related to the extremely important inflationary epoch of our universe.
Therefore, a systematically study of the supersymmetric Minkowski vacuum distribution is also very important.

\section{Conclusions}

We have shown that Starobinsky-like inflation can be simply realized as a new type of helical phase
inflation, with predictions that are in perfect agreement with new observations from Planck.
The advantages of helical phase inflation are inherited in this new model.
The so-called $\eta$-problem is directly solved by the global
$U(1)$ symmetry built into the \kah potential of $N=1$ minimal supergravity.
Helical phase inflation is driven by the phase component of a complex field,
and super-Planckian field excursions are realized within a helical phase rotation,
whereas physical fields avoid evolution into the super-Planckian regime where quantum gravitational
effects are likely to upset the slow-roll criteria.
For Starobinsky-like inflation, even though the tensor-to-scalar ratio $r$ is
generically rather small ($r\leqslant0.01$), the inflaton excursion remains
significantly larger the Planck mass during inflation, implying that the treatment of
quantum gravitational effects in a suitable UV-completion remains an open problem.
This situation is avoided in helical phase
inflation with a deformed helicoid potential, where excursions in the radial field direction are
strongly stabilized, and evolution proceeds along the trajectory of an extended flat helical minimum.
Moreover, as a PNGB, the inflaton is not expected to admit higher polynomial corrections at all.

We have identified an interpolation wherein
the helicoid potential can be continuously deformed from natural inflation parameterized by phase/axion decay
constant to Starobinsky-like inflation parameterized by $\alpha$-parameter.  In helical phase
inflation, the interpolation is uniquely characterized by a complex-valued index $\chi$. The helical
phase inflation model equipped with index $\chi$ exhibits unexpectedly rich inflationary predictions. The
full region of the $n_s-r$ plane favored by recent Planck observations can be characterized by variation of
the index $\chi$ with a fixed e-folding number $N$.

In the supergravity helical phase inflation construction, the flat phase potential is provided by phase monodromy in
the superpotential. The phase monodromy and associated pole at the field-space origin indicate that the superpotential
is an effective theory obtained after integrating out heavy fields.
We have studied the realization of such phase monodromy based on race-track non-perturbative couplings
from D-brane instanton effects. The phase monodromy relates to a global $U(1)$ symmetry in the superpotential
that is broken at the inflation scale. The \kah moduli appearing in non-perturbative terms play an important
role in generation of the complex-valued index $\chi$ for helical phase inflation. In turn, they emerge
in string moduli stabilization from non-geometric flux compactification.

The index $\chi$ has an interesting geometrical origin. In order to integrate out heavy fields, a linear
constraint is required that relates two \kah moduli appearing in the race-track non-perturbative couplings.
Constraint on the \kah moduli may be realized by the assumption of isotropy ($T_i=T$), or from
type \MyRoman{2}A orientifold compactification with NSNS, RR and geometric fluxes.
Such constraints have real coefficients and lead to
natural inflation. More general constraints with complex coefficients can be obtained by turning on
non-geometric fluxes, which generate the requisite higher-order perturbative couplings among \kah moduli and
complex structure moduli. The index $\chi$ is fixed by the vacuum expectation values of complex structure
moduli and flux quanta.  We suggest a systematic study on the
distribution of supersymmetric Minkowski vacua from non-geometric flux compactification in order to
obtain viable discrete values of the index $\chi$.
This could potentially provide more precise predictions for inflationary observables.
Conversely, it could be used to extrapolate geometric structure of the
internal space from experimental observations of the inflationary epoch.

\begin{acknowledgments}
We are grateful to J. Walker for valuable comments on the manuscript.
The work of DVN is supported in part
by the DOE grant DE-FG03-95-ER-40917. The work of TL is supported in part by
    by the Natural Science
Foundation of China under grant numbers 11135003, 11275246, and 11475238.
\end{acknowledgments}


\begin{thebibliography}{99}
\bibitem{Staro}
A.~A.~Starobinsky,
Phys.\ Lett.\ B {\bf 91}, 99 (1980).

\bibitem{oldinf}
D.~Kazanas,
Astrophys.\ J.\ {\bf 241} L59 (1980);
K.~Sato, Mon.\ Not.\ R.\ Astron.\ Soc. {\bf 195}, 467 (1981);
Phys.\ Lett.\ {\bf 99B}, 66 (1981);
A.~H.~Guth,
Phys.\ Rev.\ D {\bf 23}, 347 (1981);
A.~D.~Linde,
Phys.\ Lett.\ B {\bf 108}, 389 (1982);
A.~Albrecht and P.~Steinhardt,
Phys.~Rev.~Lett. {\bf 48}, 1220 (1982).

\bibitem{Lyth:1996im}
  D.~H.~Lyth,
  Phys.\ Rev.\ Lett.\  {\bf 78}, 1861 (1997)
  [hep-ph/9606387].
\bibitem{Li:2014vpa}
  T.~Li, Z.~Li and D.~V.~Nanopoulos,
  Phys.\ Rev.\ D {\bf 91}, no. 6, 061303 (2015)
  [arXiv:1409.3267 [hep-th]].
\bibitem{Li:2014unh}
  T.~Li, Z.~Li and D.~V.~Nanopoulos,
  arXiv:1412.5093 [hep-th].
\bibitem{Cremmer:1983bf}
  E.~Cremmer, S.~Ferrara, C.~Kounnas and D.~V.~Nanopoulos,
  Phys.\ Lett.\  B {\bf 133}, 61 (1983);
J.~R.~Ellis, A.~B.~Lahanas, D.~V.~Nanopoulos and K.~Tamvakis,
  Phys.\ Lett.\  B {\bf 134}, 429 (1984).
\bibitem{nsp}
J.~R.~Ellis, C.~Kounnas and D.~V.~Nanopoulos,
  Nucl.\ Phys.\  B {\bf 241}, 406 (1984);
  Nucl.\ Phys.\  B {\bf 247}, 373 (1984);
A.~B.~Lahanas and D.~V.~Nanopoulos,
  Phys.\ Rept.\  {\bf 145}, 1 (1987).
\bibitem{Gaillard:1995az}
  M.~K.~Gaillard, H.~Murayama and K.~A.~Olive,
  Phys.\ Lett.\ B {\bf 355}, 71 (1995)
  [hep-ph/9504307].
\bibitem{Antusch:2008pn}
  S.~Antusch, M.~Bastero-Gil, K.~Dutta, S.~F.~King and P.~M.~Kostka,
  JCAP {\bf 0901}, 040 (2009)
  [arXiv:0808.2425 [hep-ph]].
\bibitem{Kawasaki:2000yn}
  M.~Kawasaki, M.~Yamaguchi and T.~Yanagida,
  Phys.\ Rev.\ Lett.\  {\bf 85}, 3572 (2000)
  [hep-ph/0004243].
\bibitem{Li:2013nfa}
  T.~Li, Z.~Li and D.~V.~Nanopoulos,
  JCAP {\bf 1402}, 028 (2014)
  [arXiv:1311.6770 [hep-ph]].
\bibitem{Harigaya:2014qza}
  K.~Harigaya and T.~T.~Yanagida,
  Phys.\ Lett.\ B {\bf 734}, 13 (2014)
  [arXiv:1403.4729 [hep-ph]].
\bibitem{Li:2015mwa}
  T.~Li, Z.~Li and D.~V.~Nanopoulos,
  arXiv:1502.05005 [hep-ph].
\bibitem{Freese:1990rb}
  K.~Freese, J.~A.~Frieman and A.~V.~Olinto,
  Phys.\ Rev.\ Lett.\  {\bf 65}, 3233 (1990);
   F.~C.~Adams, J.~R.~Bond, K.~Freese, J.~A.~Frieman and A.~V.~Olinto,
  Phys.\ Rev.\ D {\bf 47}, 426 (1993)
  [hep-ph/9207245].
\bibitem{McAllister:2008hb}
   E.~Silverstein and A.~Westphal,
  Phys.\ Rev.\ D {\bf 78}, 106003 (2008)
  [arXiv:0803.3085 [hep-th]];
  L.~McAllister, E.~Silverstein and A.~Westphal,
  Phys.\ Rev.\ D {\bf 82}, 046003 (2010)
  [arXiv:0808.0706 [hep-th]].
\bibitem{Kim:2004rp}
  J.~E.~Kim, H.~P.~Nilles and M.~Peloso,
  JCAP {\bf 0501}, 005 (2005)
  [hep-ph/0409138].
\bibitem{Choi:2014rja}
  K.~Choi, H.~Kim and S.~Yun,
  Phys.\ Rev.\ D {\bf 90}, 023545 (2014)
  [arXiv:1404.6209 [hep-th]].

\bibitem{Tye:2014tja}
  S.-H.~H.~Tye and S.~S.~C.~Wong,
  arXiv:1404.6988 [astro-ph.CO].

\bibitem{Kappl:2014lra}
  R.~Kappl, S.~Krippendorf and H.~P.~Nilles,
  Phys.\ Lett.\ B {\bf 737}, 124 (2014)
  [arXiv:1404.7127 [hep-th]].

\bibitem{McDonald:2014oza}
  J.~McDonald,
  JCAP {\bf 1409}, no. 09, 027 (2014)
  [arXiv:1404.4620 [hep-ph]];
\bibitem{McDonald:2014nqa}
  J.~McDonald,
  JCAP {\bf 1501}, no. 01, 018 (2015)
  [arXiv:1407.7471 [hep-ph]].
\bibitem{Carone:2014cta}
  C.~D.~Carone, J.~Erlich, A.~Sensharma and Z.~Wang,
  Phys.\ Rev.\ D {\bf 91}, no. 4, 043512 (2015)
  [arXiv:1410.2593 [hep-ph]].
\bibitem{Barenboim:2014vea}
  G.~Barenboim and W.~I.~Park,
  Phys.\ Lett.\ B {\bf 741}, 252 (2015)
  [arXiv:1412.2724 [hep-ph]].
\bibitem{McDonald:2014rha}
  J.~McDonald,
  JCAP {\bf 1505}, no. 05, 014 (2015)
  [arXiv:1412.6943 [hep-ph]].
\bibitem{Barenboim:2015zka}
  G.~Barenboim and W.~I.~Park,
  Phys.\ Rev.\ D {\bf 91}, no. 6, 063511 (2015)
  [arXiv:1501.00484 [hep-ph]].
\bibitem{Achucarro:2015rfa}
  A.~Achucarro, V.~Atal and Y.~Welling,
  arXiv:1503.07486 [astro-ph.CO].
\bibitem{Blanco-Pillado:2015bha}
  J.~J.~Blanco-Pillado, M.~Dias, J.~Frazer and K.~Sousa,
  arXiv:1503.07579 [astro-ph.CO].
\bibitem{Barenboim:2015lla}
  G.~Barenboim and W.~I.~Park,
  arXiv:1504.02080 [astro-ph.CO].
\bibitem{Peloso:2015dsa}
  M.~Peloso and C.~Unal,
  JCAP {\bf 1506}, no. 06, 040 (2015)
  [arXiv:1504.02784 [astro-ph.CO]].

\bibitem{Planck:2015xua}
  P.~A.~R.~Ade {\it et al.}  [Planck Collaboration],
  arXiv:1502.01589 [astro-ph.CO].
\bibitem{Ade:2015oja}
  P.~A.~R.~Ade {\it et al.}  [Planck Collaboration],
  arXiv:1502.02114 [astro-ph.CO].
\bibitem{Ade:2015tva}
  P.~A.~R.~Ade {\it et al.}  [BICEP2 and Planck Collaborations],
  [arXiv:1502.00612 [astro-ph.CO]].

\bibitem{Ellis:2013xoa}
  J.~Ellis, D.~V.~Nanopoulos and K.~A.~Olive,
  Phys.\ Rev.\ Lett.\  {\bf 111}, 111301 (2013)
  [Phys.\ Rev.\ Lett.\  {\bf 111}, no. 12, 129902 (2013)]
  [arXiv:1305.1247 [hep-th]];
\bibitem{Ellis:2013nxa}
  J.~Ellis, D.~V.~Nanopoulos and K.~A.~Olive,
  JCAP {\bf 1310}, 009 (2013)
  [arXiv:1307.3537].

\bibitem{Ferrara:2013rsa}
  S.~Ferrara, R.~Kallosh, A.~Linde and M.~Porrati,
  Phys.\ Rev.\ D {\bf 88}, no. 8, 085038 (2013)
  [arXiv:1307.7696 [hep-th]].
\bibitem{Kallosh:2013yoa}
  R.~Kallosh, A.~Linde and D.~Roest,
  JHEP {\bf 1311}, 198 (2013)
  [arXiv:1311.0472 [hep-th]].

\bibitem{Ellis:2014cma}
  J.~Ellis, N.~E.~Mavromatos and D.~V.~Nanopoulos,
  Phys.\ Lett.\ B {\bf 732}, 380 (2014)
  [arXiv:1402.5075 [hep-th]].
\bibitem{Ellis:2014gxa}
  J.~Ellis, M.~A.~G.~Garcia, D.~V.~Nanopoulos and K.~A.~Olive,
  JCAP {\bf 1408}, 044 (2014)
  [arXiv:1405.0271 [hep-ph]].
\bibitem{Kounnas:2014gda}
  C.~Kounnas, D.~Lüst and N.~Toumbas,
  Fortsch.\ Phys.\  {\bf 63}, 12 (2015)
  [arXiv:1409.7076 [hep-th]].
\bibitem{Ellis:2014opa}
  J.~Ellis, M.~A.~G.~García, D.~V.~Nanopoulos and K.~A.~Olive,
  JCAP {\bf 1501}, no. 01, 010 (2015)
  [arXiv:1409.8197 [hep-ph]].
\bibitem{Higaki:2015kta}
  T.~Higaki and F.~Takahashi,
  JHEP {\bf 1503}, 129 (2015)
  [arXiv:1501.02354 [hep-ph]].
\bibitem{Kannike:2015apa}
  K.~Kannike, G.~Hütsi, L.~Pizza, A.~Racioppi, M.~Raidal, A.~Salvio and A.~Strumia,
  JHEP {\bf 1505}, 065 (2015)
  [arXiv:1502.01334 [astro-ph.CO]].
\bibitem{Ozkan:2015iva}
  M.~Ozkan, Y.~Pang and S.~Tsujikawa,
  arXiv:1502.06341 [astro-ph.CO].
\bibitem{Ellis:2015xna}
  J.~Ellis, M.~A.~G.~Garcia, D.~V.~Nanopoulos and K.~A.~Olive,
  arXiv:1507.02308 [hep-ph].
\bibitem{Cicoli:2008gp}
  M.~Cicoli, C.~P.~Burgess and F.~Quevedo,
  JCAP {\bf 0903}, 013 (2009)
  [arXiv:0808.0691 [hep-th]].
\bibitem{Burgess:2013sla}
  C.~P.~Burgess, M.~Cicoli and F.~Quevedo,
  JCAP {\bf 1311}, 003 (2013)
  [arXiv:1306.3512 [hep-th]].
\bibitem{Cicoli:2013oba}
  M.~Cicoli, S.~Downes and B.~Dutta,
  JCAP {\bf 1312}, 007 (2013)
  [arXiv:1309.3412 [hep-th]].
\bibitem{Blumenhagen:2015qda}
  R.~Blumenhagen, A.~Font, M.~Fuchs, D.~Herschmann and E.~Plauschinn,
  Phys.\ Lett.\ B {\bf 746}, 217 (2015)
  [arXiv:1503.01607 [hep-th]].

\bibitem{Kachru:2002sk}
  S.~Kachru, M.~B.~Schulz, P.~K.~Tripathy and S.~P.~Trivedi,
  JHEP {\bf 0303}, 061 (2003)
  [hep-th/0211182].
\bibitem{Shelton:2005cf}
  J.~Shelton, W.~Taylor and B.~Wecht,
  JHEP {\bf 0510}, 085 (2005)
  [hep-th/0508133].

\bibitem{Aldazabal:2006up}
  G.~Aldazabal, P.~G.~Camara, A.~Font and L.~E.~Ibanez,
  JHEP {\bf 0605}, 070 (2006)
  [hep-th/0602089].
\bibitem{Shelton:2006fd}
  J.~Shelton, W.~Taylor and B.~Wecht,
  JHEP {\bf 0702}, 095 (2007)
  [hep-th/0607015].

\bibitem{Micu:2007rd}
  A.~Micu, E.~Palti and G.~Tasinato,
  JHEP {\bf 0703}, 104 (2007)
  [hep-th/0701173].
\bibitem{Font:2008vd}
  A.~Font, A.~Guarino and J.~M.~Moreno,
  JHEP {\bf 0812}, 050 (2008)
  [arXiv:0809.3748 [hep-th]].
\bibitem{Guarino:2008ik}
  A.~Guarino and G.~J.~Weatherill,
  JHEP {\bf 0902}, 042 (2009)
  [arXiv:0811.2190 [hep-th]].
\bibitem{deCarlos:2009qm}
  B.~de Carlos, A.~Guarino and J.~M.~Moreno,
  JHEP {\bf 1002}, 076 (2010)
  [arXiv:0911.2876 [hep-th]].
\bibitem{Aldazabal:2011yz}
  G.~Aldazabal, D.~Marques, C.~Nunez and J.~A.~Rosabal,
  Nucl.\ Phys.\ B {\bf 849}, 80 (2011)
  [arXiv:1101.5954 [hep-th]].
\bibitem{Hassler:2014mla}
  F.~Hassler, D.~Lust and S.~Massai,
  arXiv:1405.2325 [hep-th].
\bibitem{Blumenhagen:2015kja}
  R.~Blumenhagen, A.~Font, M.~Fuchs, D.~Herschmann, E.~Plauschinn, Y.~Sekiguchi and F.~Wolf,
  Nucl.\ Phys.\ B {\bf 897}, 500 (2015)
  [arXiv:1503.07634 [hep-th]].
\bibitem{Ketov:2015tpa}
  S.~V.~Ketov and T.~Terada,
  arXiv:1509.00953 [hep-th].
\bibitem{Andriot:2014qla}
  D.~Andriot and A.~Betz,
  JHEP {\bf 1504}, 006 (2015)
  [arXiv:1411.6640 [hep-th]].


\bibitem{Ibanez:2012zz}
  L.~E.~Ibanez and A.~M.~Uranga,
  Cambridge, UK: Univ. Pr. (2012) 673 p

\bibitem{Gukov:1999ya}
  S.~Gukov, C.~Vafa and E.~Witten,
  Nucl.\ Phys.\ B {\bf 584}, 69 (2000)
  [Nucl.\ Phys.\ B {\bf 608}, 477 (2001)]
  [hep-th/9906070].

\bibitem{Robbins:2007yv}
  D.~Robbins and T.~Wrase,
  JHEP {\bf 0712}, 058 (2007)
  [arXiv:0709.2186 [hep-th]].
\bibitem{Shukla:2015rua}
  P.~Shukla,
  arXiv:1505.00544 [hep-th].
\bibitem{Shukla:2015bca}
  P.~Shukla,
  arXiv:1507.01612 [hep-th].




\bibitem{Gray:2008zs}
  J.~Gray, Y.~H.~He, A.~Ilderton and A.~Lukas,
  Comput.\ Phys.\ Commun.\  {\bf 180}, 107 (2009)
  [arXiv:0801.1508 [hep-th]].
\bibitem{Kachru:2002he}
  S.~Kachru, M.~B.~Schulz and S.~Trivedi,
  JHEP {\bf 0310}, 007 (2003)
  [hep-th/0201028].
\bibitem{DeWolfe:2004ns}
  O.~DeWolfe, A.~Giryavets, S.~Kachru and W.~Taylor,
  JHEP {\bf 0502}, 037 (2005)
  [hep-th/0411061].
\end{thebibliography}
\end{document}